\documentclass[12pt,superscriptaddress]{article}
\pdfoutput=1
\usepackage{jheppub}
\usepackage{graphicx}  
\usepackage{dcolumn}   
\usepackage{bm}        
\usepackage{amssymb}   

\newcommand{\beq}{\begin{equation}}
\newcommand{\eeq}[1]{\label{#1}\end{equation}}
\def\beqa{\begin{eqnarray}}
\def\eeqa#1{\label{#1}\end{eqnarray}}
\newcommand{\eeqn}{\end{equation}}
\newcommand{\CR}{\notag \\}
\newcommand{\leqn}[1]{(\ref{#1})}                                                                        
                       
\newcommand{\bspace}{\!\!\!\!}
\newcommand{\me}{\ensuremath{E\bspace /}~}
\newcommand{\met}{\ensuremath{E\bspace /_T}}

\def\stacksymbols #1#2#3#4{\def\theguybelow{#2}
    \def\vp{\lower#3pt}
    \def\sp{\baselineskip0pt\lineskip#4pt}
    \mathrel{\mathpalette\intermediary#1}}

\def\intermediary#1#2{\vp\vbox{\sp
     \everycr={}\tabskip0pt
     \halign{$\mathsurround0pt#1\hfil##\hfil$\crcr#2\crcr
              \theguybelow\crcr}}}

\begin{document}


\title{Dark Matter Search at a Linear Collider: Effective Operator Approach}

\author{Yoonseok Chae} 
\author{and Maxim Perelstein}

\affiliation{Laboratory of Elementary Particle Physics, 
	     Cornell University, Ithaca, NY 14853, USA}
	     
\emailAdd{yc654@cornell.edu} \emailAdd{mp325@cornell.edu}     

\abstract{Experiments at electron-positron colliders can search for dark matter particle pair-production in association with a photon. We estimate the sensitivity of this search at the proposed International Linear Collider (ILC), under a variety of run scenarios. We employ the effective operator formalism to provide a quasi-model-independent theoretical description of the signal, and present the reach of the ILC in terms of the scale $\Lambda$ suppressing the dark matter-electron coupling operator. We find that at the 250 GeV center-of-mass energy, the ILC can probe $\Lambda$ up to $1-1.2$ TeV, a factor of $2.5-3$ above the best current bounds from LEP-2. With 1 TeV energy and polarized beams, the reach can be extended to $3-4$ TeV. The ILC can discover this signature even if annihilation to electrons provides only a small fraction of the total dark matter annihilation rate in the early universe. We also argue that large regions of parameter space allowed by current LHC and direct detection bounds will be accessible at the ILC.  
}


\maketitle

\newpage

\section{Introduction}

While the existence of dark matter has been firmly established through observation of its gravitational effects on multiple scales, its microscopic nature remains unknown. It is clear that none of the Standard Model (SM) particles can account for dark matter, so that new physics must be involved. This situation motivated a variety of searches for dark matter through non-gravitational interactions, such as direct, indirect and collider signatures. No firm detection has been achieved as yet, but several intriguing hints have been reported. 

The focus of this paper is on collider signatures of dark matter. From the theoretical point of view, a very attractive class of dark matter candidates is the weakly interacting massive particles, or WIMPs. WIMPs fit naturally in a variety of popular extensions of the SM at the electroweak scale, and they generally have the right order-of-magnitude thermal relic density to fit the observed dark matter abundance (the so-called ``WIMP miracle"). It is plausible that WIMPs could be produced at significant rates at current or near-future high-energy colliders, giving potentially observable experimental signatures. Extensive studies of such signatures appeared in the literature. However, most studies are performed in the context of a particular particle physics model, such as supersymmetry (usually with very specific superpartner spectrum in mind), Little Higgs, etc. Given that a large number of particle physics models contain WIMPs, it is highly desirable to identify and focus on signatures which do not depend sensitively on model details, but are as model-independent as possible. 

The first example of a model-independent approach to collider signatures of dark matter appeared in 2004 in Ref.~\cite{BMP}. This paper argued that the process of pair-creation of dark matter particles in, for example, $e^+e^-$ collisions, is just the reverse of one of the annihilation processes which determine the thermal relic density of the WIMPs. This allows one to use the measured abundance of dark matter to predict the rate of its production at an $e^+e^-$ collider. An observable signature is obtained if a photon is emitted before the collision: since the WIMPs escape the detector unobserved, the signature is $\gamma+\me$. Using soft/collinear factorization theorems, the rate of this process can also be predicted in a model-independent way, up to a small number of unknown phenomenological parameters, each with a clear physical significance. Detailed Monte Carlo studies of the sensitivity of the proposed International Linear Collider (ILC) to this process have since been performed~\cite{ListEtAl1,ListEtAl2}.       

More recently, an alternative theoretical framework for a quasi-model-independent studies of dark matter at colliders was proposed~\cite{operators}. The starting point is a Lagrangian in which the dark matter is coupled to the SM through a set of effective (non-renormalizable) operators. Such operators represent the effect of heavy ``mediator" particles, but all dependence on the specifics of the mediators is contained in a small number of operator coefficients. Most of the applications of this approach have focused on dark matter signatures at hadron colliders, $j+\met$ and $\gamma+\met$. 
These are induced by the same operators as the dark matter signal in direct detection experiments, so correlations between the two types of experiments can be understood. Tevatron and LHC experiments have already reported bounds from searches performed with this approach~\cite{exp}. Effective operator formalism can also be applied to dark matter production in $e^+e^-$ collisions. This has been done in Ref.~\cite{LEPshines}, which reported bounds on dark matter-SM couplings from non-observation of anomalous  $\gamma+\me$ events at LEP-2. 

An energy frontier $e^+e^-$ collider, the International Linear Collider (ILC), is one of the most compelling proposals for future major facilities in fundamental physics. The most immediate physics motivation for such a machine is to study the recently discovered 125 GeV boson in detail, and confirm or disprove its identification with the SM Higgs. For this task, the center-of-mass energy of $e^+e^-$ collisions of about 250 GeV would be ideal. In the future, the accelerator can be upgraded to reach the collision energies of $1-1.5$ TeV. One of the most intriguing experiments that can be performed at the ILC is a search for dark matter. Our goal in this paper is to evaluate the potential sensitivity of such a search, under a variety of machine parameters suggested by recent design studies. To minimize dependence on unknown particle physics, we will employ the effective operator formalism mentioned above.

Given the time scale for the completion of design and construction of the next $e^+e^-$ collider, it is quite possible that 
WIMP dark matter will be discovered, for example at the LHC or via direct detection, before the experiments discussed in this paper can begin. In this scenario, the main task of the $e^+e^-$ collider will be to perform detailed measurements of the dark matter particle properties. Still, the dark matter signal will need to be established first, so our study would still be relevant. Moreover, it has been seen in similar examples that the reach for discovery and model discrimination is typically roughly similar (see {\it e.g.}~\cite{KKMP,PP}), so our study gives at least a rough sense of how well different models (for example, different effective operators for SM-WIMP coupling) can be distinguished.  

The rest of the paper is organized as follows: We describe the theoretical framework of the analysis, the collider run scenarios considered, and other assumptions, in Sec.~\ref{sec:setup}. Our estimates of the dark matter search reach at the ILC are presented and discussed in Sec.~\ref{sec:results}. We summarize the main conclusions of the analysis in Sec.~\ref{sec:conc}. Appendix~\ref{app:analytics} contains the analytic formulas for the differential cross sections of the radiative WIMP production in (polarized) $e^+e^-$ collisions, which have not previously appeared in the literature.

\section{Setup}
\label{sec:setup}

\begin{figure}[tb]
\begin{center}
\centerline {
\includegraphics{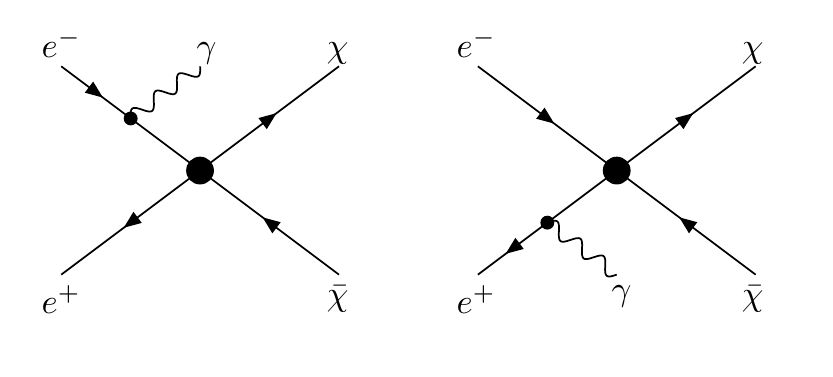}
}
\caption{Feynman diagrams for radiative WIMP pair-production in $e^+e^-$ collisions, in the operator formalism.}
\label{fig:diagrams}
\end{center}
\end{figure}

While the operator formalism can be used for WIMPs of any spin, we will assume, for concreteness, that the WIMP is a spin-1/2, Dirac fermion $\chi$. The coupling of the WIMPs to electrons and positrons has the form 
\beq
{\cal L}_{\rm int} = \frac{1}{\Lambda^2} \,{\cal O}_i\,,
\eeq{Lint}
where $\Lambda$ roughly corresponds to the energy scale of new physics that provides the coupling, and ${\cal O}_i$ is one of the following four-fermion operators~\cite{LEPshines}:
\beqa
{\cal O}_V &=& (\bar{\chi} \gamma_\mu \chi) (\bar{\ell} \gamma^\mu \ell) \,,~~~~~~~~{\rm (vector)} \CR
{\cal O}_S &=& (\bar{\chi} \chi) (\bar{\ell} \ell)\,,~~~~~~~~~~~~~~{\rm (scalar,~}s-{\rm channel)} \CR
{\cal O}_A &=& (\bar{\chi} \gamma_\mu \gamma_5 \chi) (\bar{\ell} \gamma^\mu \gamma^5 \ell) \,,~~{\rm (axial-vector)} \CR
{\cal O}_t &=& (\bar{\chi} \ell) (\bar{\ell} \chi)\,,~~~~~~~~~~~~~~{\rm (scalar,~}t{\rm -channel)} .
\eeqa{ops}
The notation in parenthesis describes the simplest kind of a mediator particle that would induce each operator. We will always consider the case when the mediator mass is well above the collision energy $\sqrt{s}$, and our results will not depend on how the operators~\leqn{ops} are induced; the names are only used as a convenient way to label operators. 
Since the WIMPs do not interact in the detector, the $2\rightarrow 2$ process $e^+e^-\to\bar{\chi}\chi$ is invisible; an extra ``tag" particle needs to be added to the final state to make it observable. A photon can always be emitted from the initial state independently of the nature of the WIMPs and their couplings, making it a robust choice for the tag particle~\cite{BMP}. We will thus consider the process $e^+e^-\to\bar{\chi}\chi\gamma$, mediated by Feynman diagrams in Fig.~\ref{fig:diagrams}, and leading to the observable $\gamma+\me$ final state. We have computed the double-differential cross sections, $\frac{d^2\sigma}{dE_\gamma d\cos\theta}$, analytically for each of the four interactions listed in~\leqn{ops} and for all possible combinations of electron and positron beam polarizations. The formulas are presented in Appendix~\ref{app:analytics}.

\begin{table}[tb]
  \centering
  \begin{tabular}{|l|l|l|l|l|l|}
    \hline
 & ILC-250 & ILC-500 & ILC-1000 & ILC-500P & ILC-1000P  \\
 \hline
$\sqrt{s}$, GeV & 250 & 500 & 1000 & 500 & 1000  \\
$L_{\rm int}$, fb$^{-1}$ & 250 & 500 & 1000 & 250 & 500  \\
$P_-/P_+$ & 0/0 & 0/0 & 0/0 & +0.8/+0.5 & +0.8/+0.5  \\
    \hline
    $\sigma_{\rm bg}$, fb & 1.07 & 1.74 & 2.36 & 0.21 & 0.25 \\
    \hline
  \end{tabular}
  \caption{Accelerator parameters for each of the run scenarios studied in this paper, and the SM background cross sections within the experimentally accessible region.}
  \label{tab:pars}
\end{table}

The main irreducible background to the search for the $\gamma+\me$ signature is the SM process $e^+e^-\to \nu\bar{\nu}\gamma$. We have evaluated the cross section for this process using {\tt MadGraph/MadEvent v5.0}~\cite{MG}. We ignore instrumental backgrounds, and assume a systematic error on the background prediction of 0.3\%. We combine the statistical and systematic errors in quadrutures to obtain the significance:
\beq
{\rm Sig} = \frac{N_{\rm sig}}{\sqrt{N_{\rm bg} + (0.003 N_{\rm bg})^2}}.
\eeq{Ns}
The reach of a given experiment is estimated by requiring Sig$\geq 3$. In all scenarios considered below, $N_{\rm bg}\gg 1$, so the use of Gaussian statistics is well justified. We present results for several sets of collider parameters, summarized in Table~\ref{tab:pars}. For the proposed ILC, we follow~\cite{Peskin} and assume three stages of operation: ILC-250 ($\sqrt{s}=250$ GeV, $L_{\rm int}=250$ fb$^{-1}$); ILC-500 ($\sqrt{s}=500$ GeV, $L_{\rm int}=500$ fb$^{-1}$); and ILC-1000 ($\sqrt{s}=1$ TeV, $L_{\rm int}=1000$ fb$^{-1}$).
We assume unpolarized beams in these three scenarios. In addition, we consider the possibility of beam polarization at 500 GeV and 1 TeV, taking the electron and positron polarizations to be $P_-=+0.8$ and $P_+=+0.5$, respectively.\footnote{Our notation for polarization is discussed in Appendix~\ref{app:analytics}. The signs are chosen so that electrons are predominantly right-handed and positrons left-handed, suppressing the contribution of the diagram with the $t$-channel $W$ exchange to the background.} (Note that in the polarized beam case, the uncertainty in the degree of polarization should be included in the systematic error; the 0.3\% error assumed in our study would require knowledge of beam polarization at this level or better.)  
On the detector side, the parameters relevant for our study are the minimum photon energy $E_\gamma^{\rm min}$, and the minimum value of the angle of the photon to the beamline, characterized by $|\cos\theta_\gamma|^{\rm max}$, for which the photon can be detected. For all scenarios, we use $E_\gamma^{\rm min}=8$ GeV and $|\cos\theta_\gamma|^{\rm max}=0.995$. In addition, to further suppress the background, we discard events with photon energies corresponding to recoil against an on-shell $Z$. Specifically, we define $z=2E_\gamma/\sqrt{s}$, and discard events with $z\in [0.8, 0.9]$ for 
ILC-250, $z\in [0.95, 0.98]$ for ILC-500(P), and $z\in [0.98, 0.99]$ for ILC-1000(P).
The cross sections of the SM background process $e^+e^-\to \nu\bar{\nu}\gamma$, after all cuts, are listed in Table~\ref{tab:pars}.

\section{Results}
\label{sec:results}

\begin{figure}[h!]
\begin{center}
\centerline {
\includegraphics[width=3.3in]{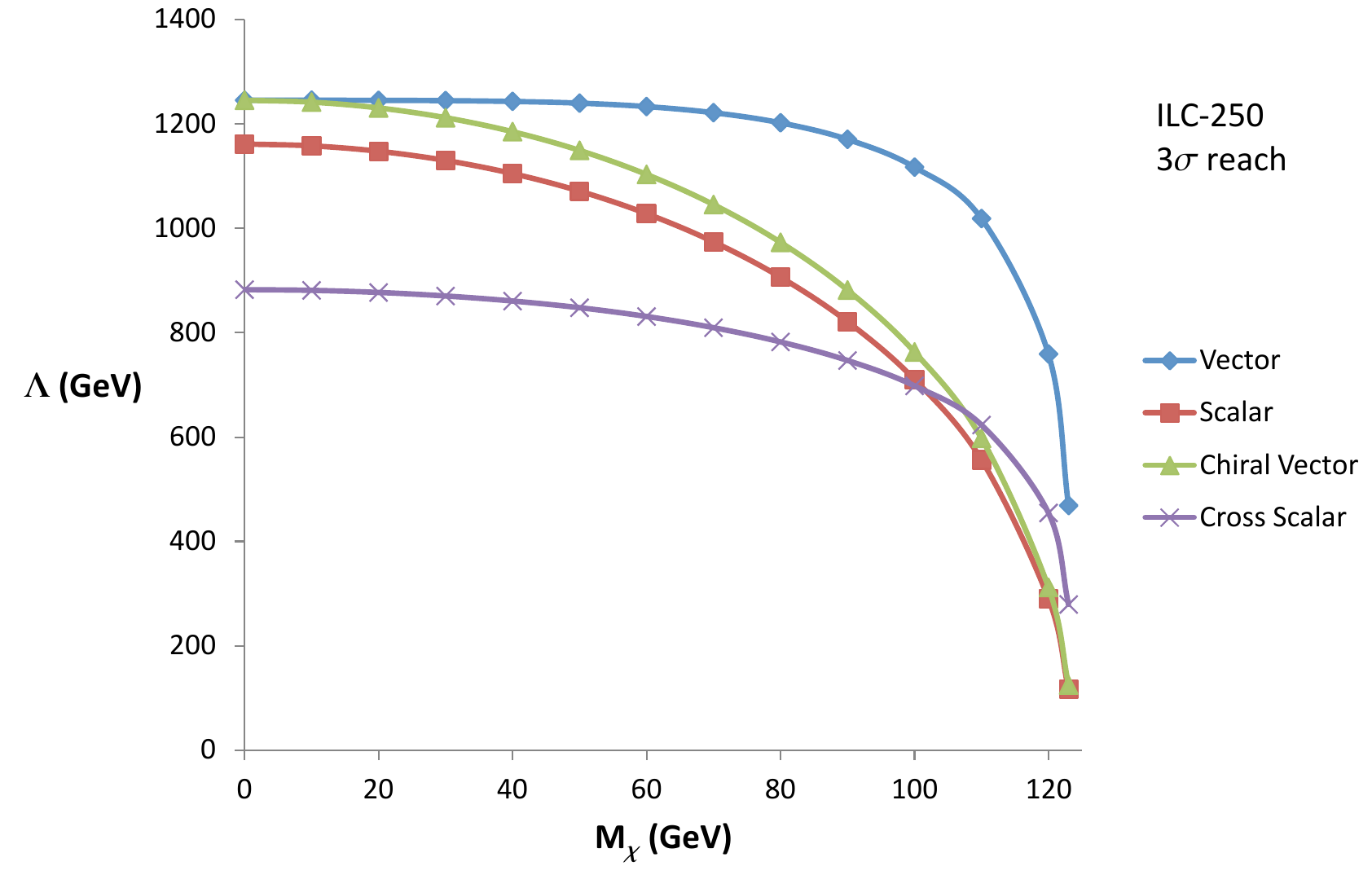}
\includegraphics[width=3.3in]{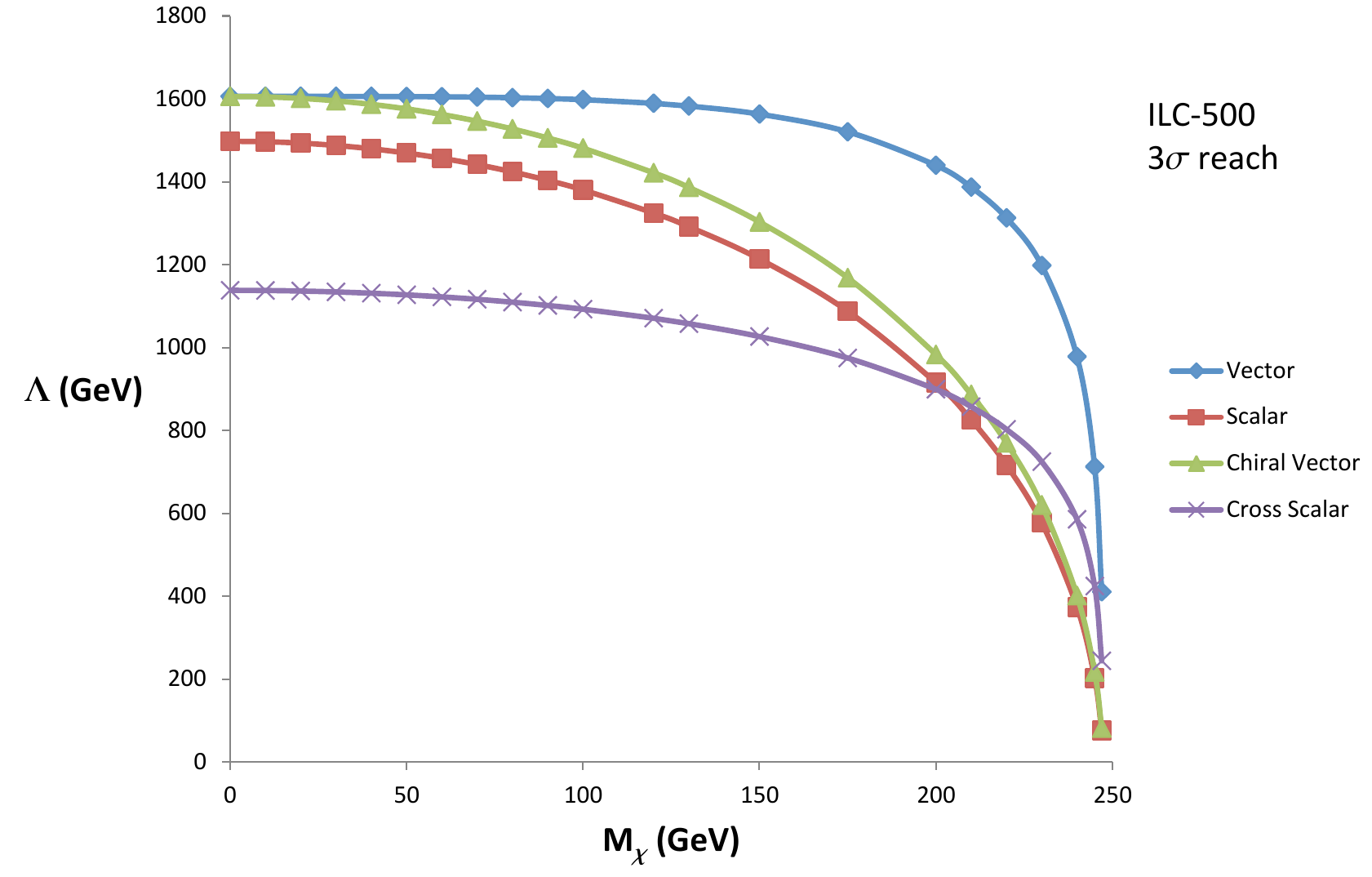}
}
\vskip.3cm
\centerline {
\includegraphics[width=3.3in]{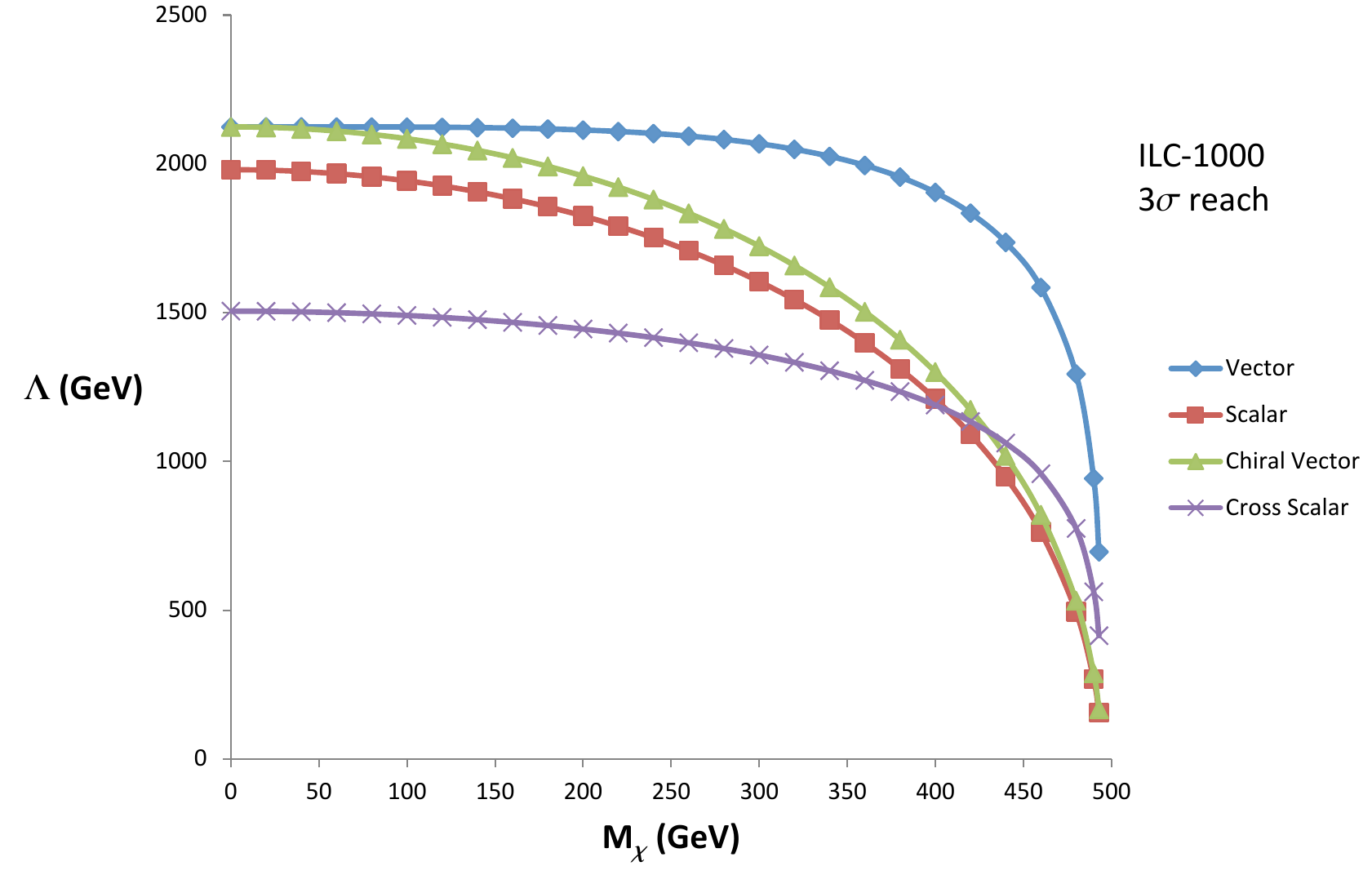}
\includegraphics[width=3.3in]{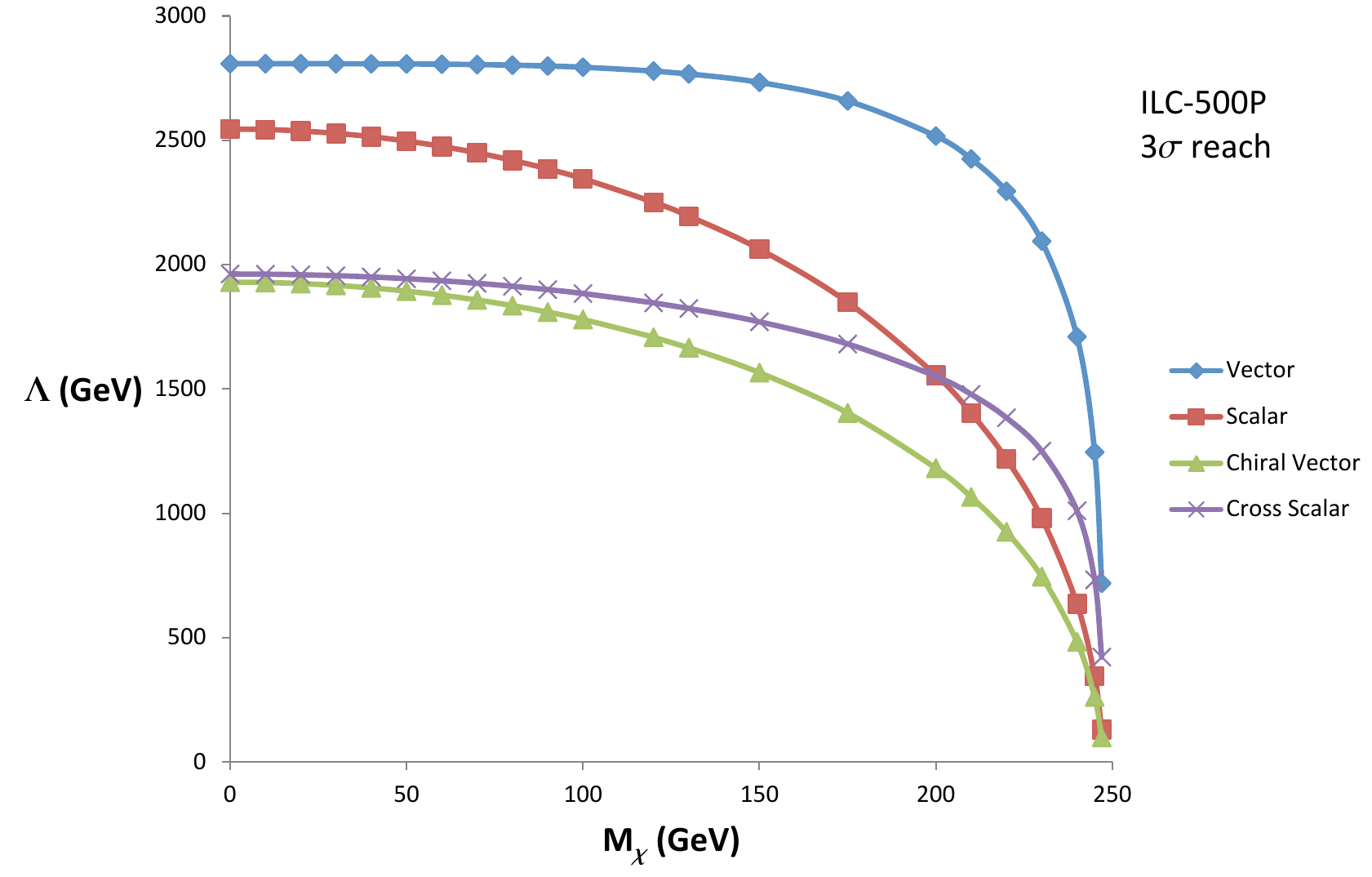}
}
\vskip.3cm
\centerline {
\includegraphics[width=3.3in]{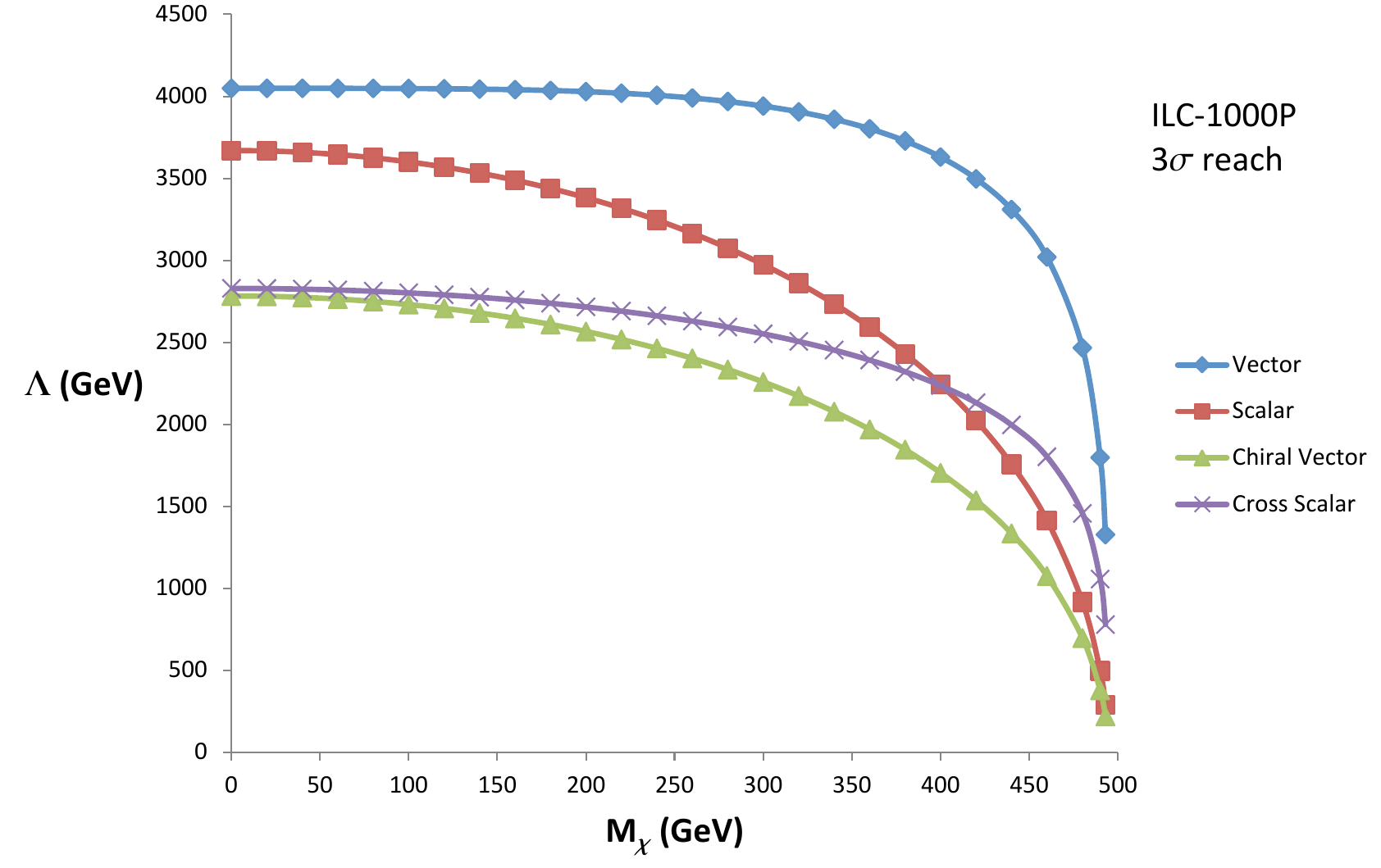}
}
\caption{Reach of the ILC dark matter searches, for the 5 run scenarios defined in Table~\ref{tab:pars}, as a function of the dark matter mass $M_\chi$ and the scale $\Lambda$ of the WIMP-electron coupling operators. The regions below the curves are accessible at the 3-sigma level.}
\label{fig:reachplots}
\end{center}
\end{figure}

Our estimates of the reach of the ILC dark matter searches, with parameters described above, are given in Fig.~\ref{fig:reachplots}. In terms of the scale $\Lambda$, the accessible regions are approximately independent of the dark matter particle mass, until the kinematic reach of the collider (roughly $M_\chi\approx \sqrt{s}/2$) is reached. The dependence on the helicity structure of the operator is significant but not overwhelming, with all bounds within roughly 30\% of each other. A 250 GeV ILC (the ``Higgs factory") is sensitive to scales $\Lambda$ up to about $1-1.2$ TeV, a factor of  
$2.5-3$ higher than the LEP bounds given in Ref.~\cite{LEPshines}. Further improvements are possible by increasing collision energy and/or using polarized beams. For example, a 1 TeV collider with polarized beams can probe $\Lambda$ up to $3-4$ TeV, depending on the operator. In all cases, the reach is significantly higher than the center-of-mass energy in $e^+e^-$ collisions, justifying the use of the effective operator formalism in our analysis. 

\begin{figure}[h!]
\begin{center}
\centerline {
\includegraphics[width=3.3in]{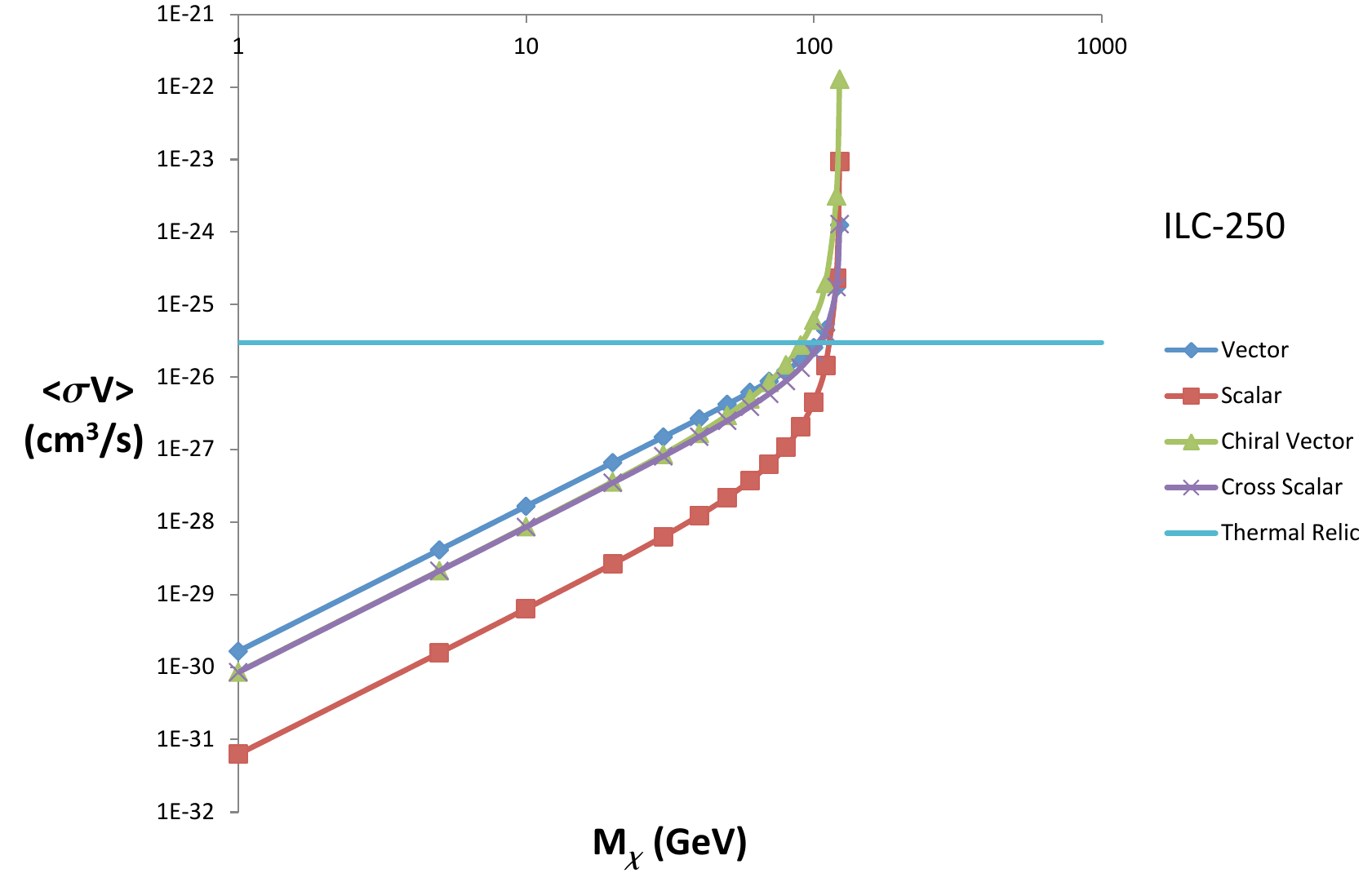}
\includegraphics[width=3.3in]{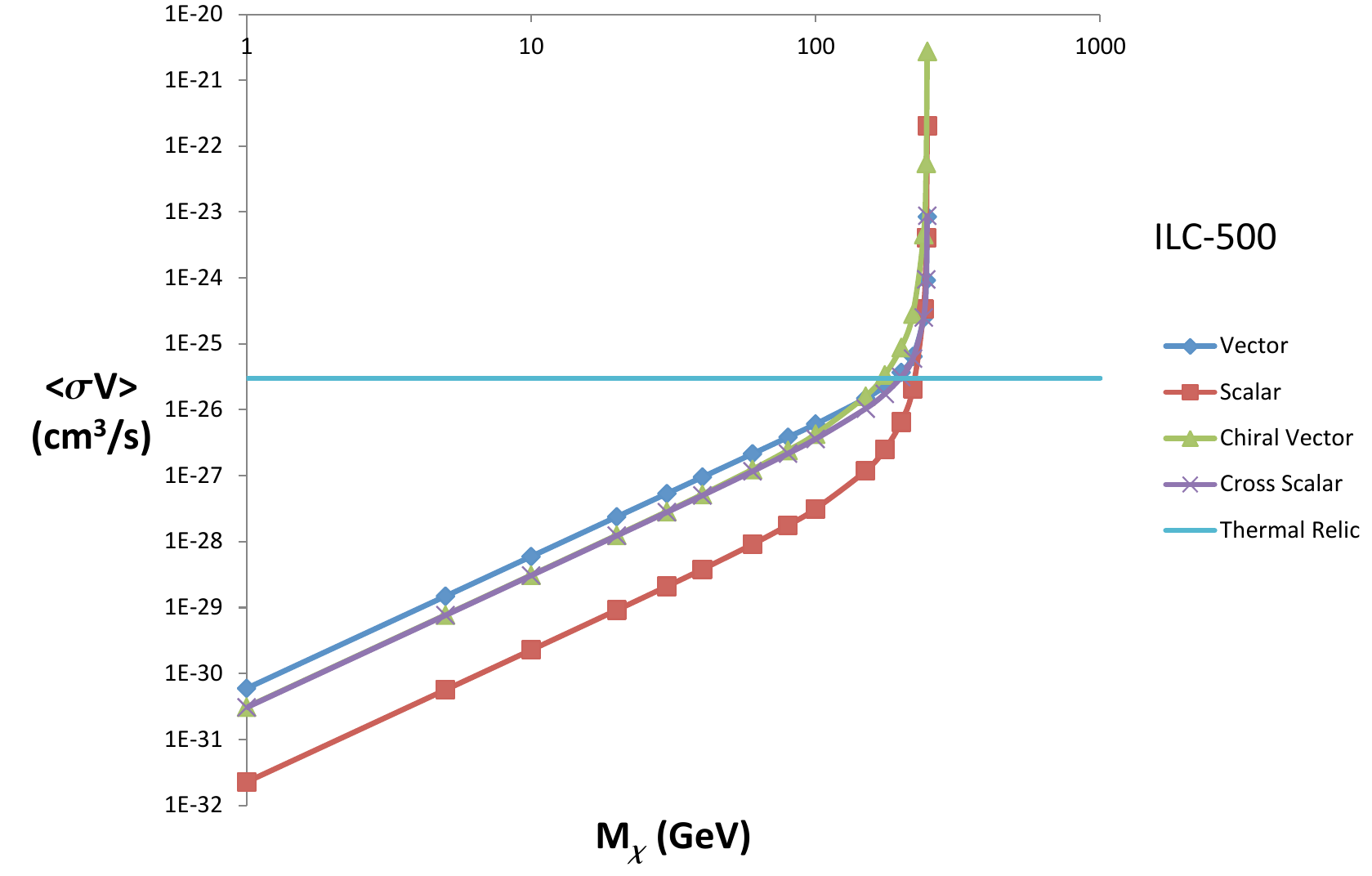}
}
\centerline {
\includegraphics[width=3.3in]{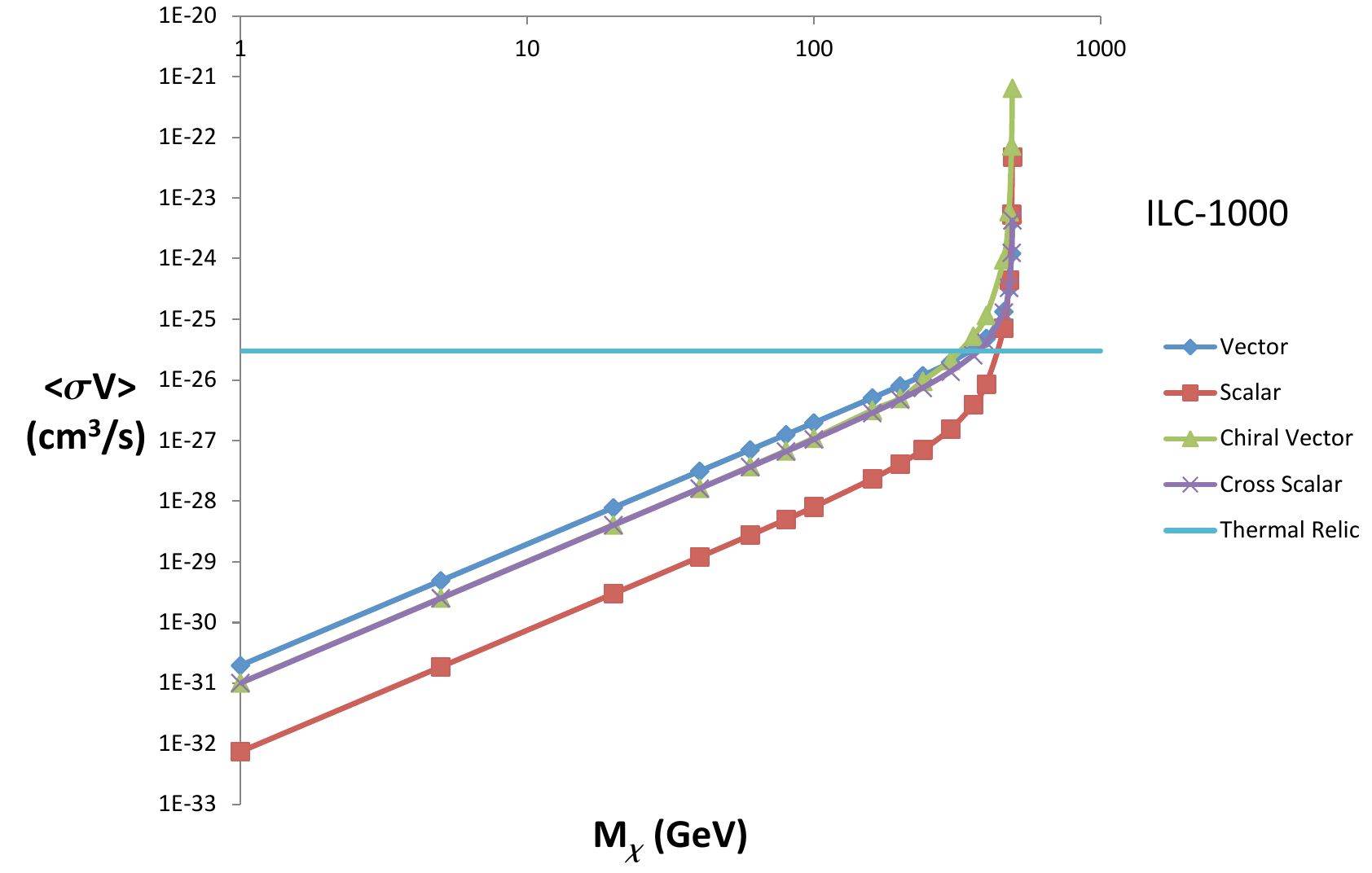}
\includegraphics[width=3.3in]{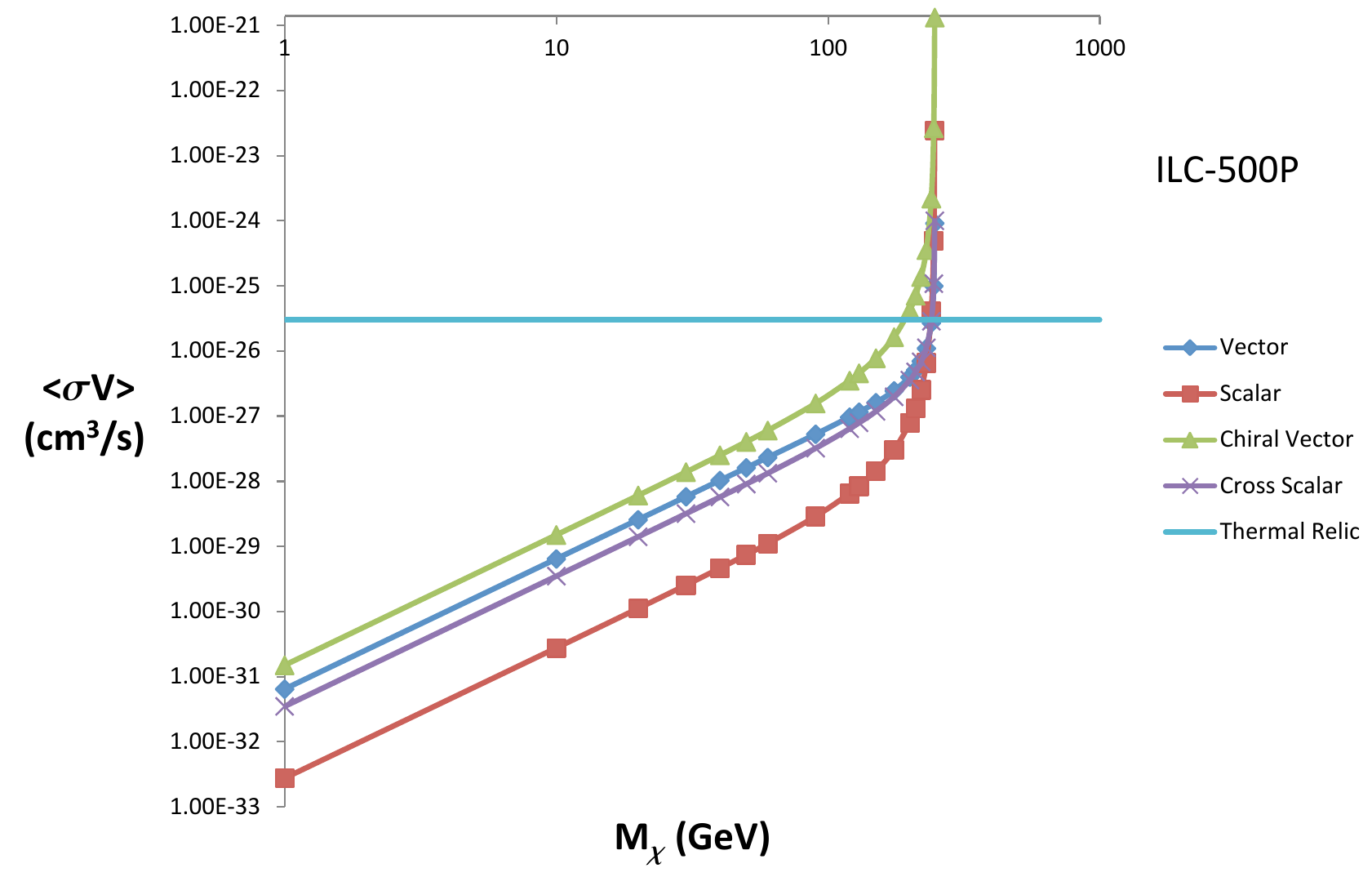}
}
\centerline {
\includegraphics[width=3.3in]{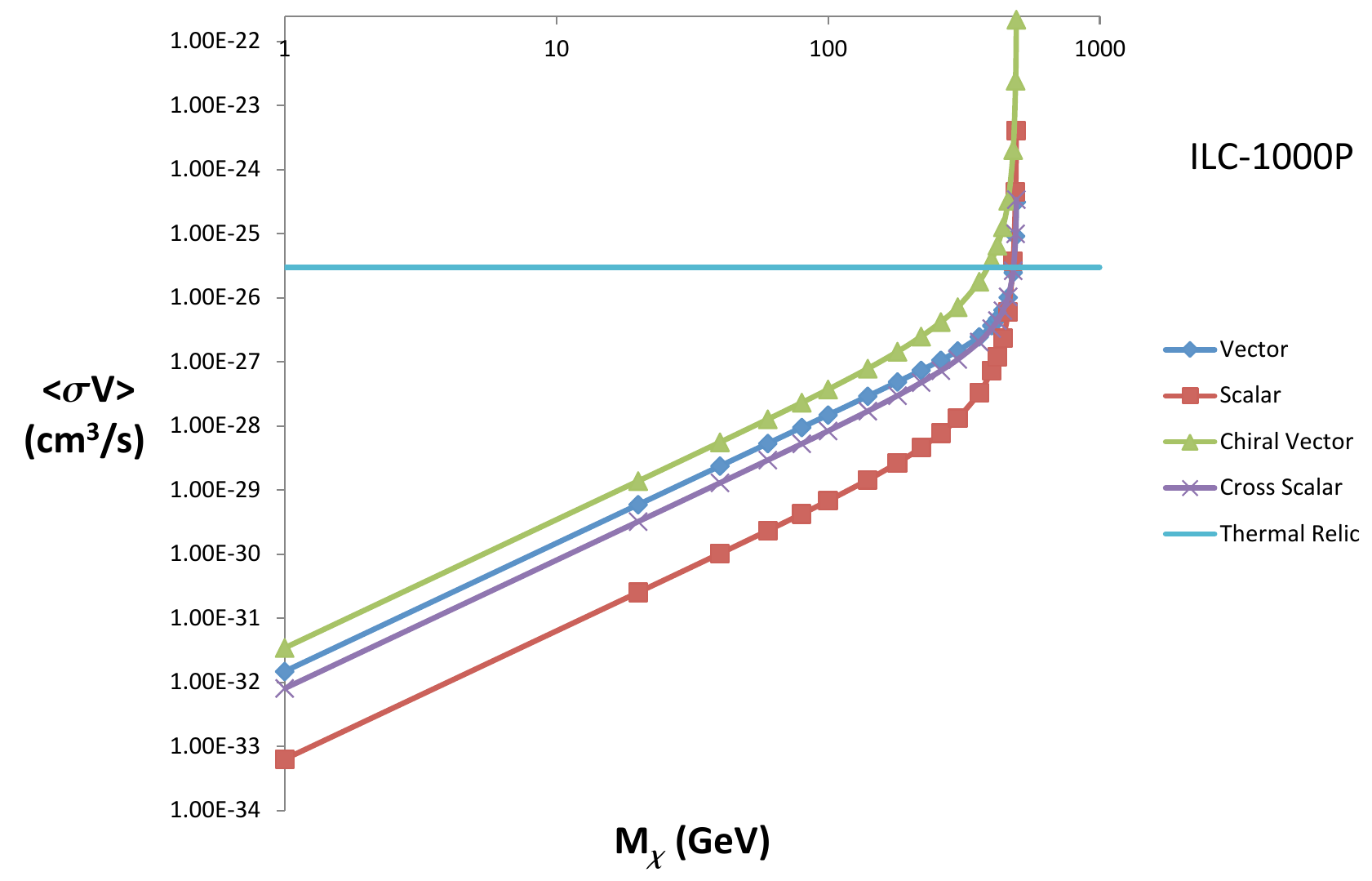}
}
\caption{Reach of the ILC dark matter searches, as a function of the dark matter mass $M_\chi$ and the thermally averaged annihilation cross section at freeze-out, $\langle \sigma (\chi\chi\to e^+e^-)v\rangle$. The regions above the curves are accessible at the 3-sigma level. The value of $\langle \sigma (\chi\chi\to e^+e^-)v\rangle$ needed to obtain the correct relic density, in the absence of other annihilation channels, is also shown.}
\label{fig:RD}
\end{center}
\end{figure}

A lower bound on $\Lambda$ is equivalent to an upper bound on the annihilation cross section $\chi\chi\to e^+e^-$. Let us assume that $\chi$ is a thermal relic, and that $e^+e^-$ is the dominant annihilation channel in the early universe. The thermally-averaged value of this cross section, evaluated at the temperature at which decoupling takes place, determines the present density of WIMPs. In Fig.~\ref{fig:RD}, we plot the projected ILC upper bounds on the quantity $\langle \sigma (\chi\chi\to e^+e^-)v\rangle$, where $v$ is the relative velocity of the colliding WIMPs and the brackets denote thermal averaging. (We assume $\langle v^2\rangle=0.24$ at freeze-out.) The observed present density of WIMPs requires $\langle \sigma (\chi\chi\to e^+e^-)v\rangle\approx 3\times 10^{-26}$ cm$^3$/sec. It is clear that the ILC will be able to probe significantly smaller cross sections, throughout the kinematically accessible WIMP mass range. In particular, in the low WIMP mass region, between 1 and 10 GeV, the ILC-250 will be able to discover the WIMPs 
even if their annihilation to electrons provides only $\sim 10^{-5}-10^{-3}$ of the critical value. The LEP bounds on the annihilation cross section~\cite{LEPshines} would be improved by one to two orders of magnitude. 

\begin{figure}[tb]
\begin{center}
\centerline {
\includegraphics{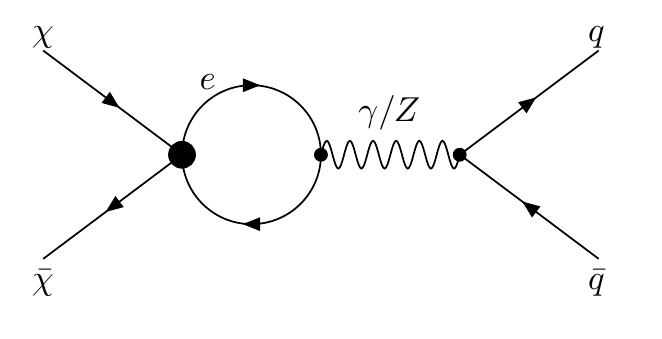}
}
\caption{The one-loop diagram inducing the WIMP-quark coupling from a tree-level WIMP-electron operator.}
\label{fig:oneloop}
\end{center}
\end{figure}

It is also interesting to compare the ILC reach with that of direct detection searches for dark matter, as well as searches at the LHC. The LHC searches for WIMP pair-production (in association with a jet or a photon) in $pp$ collisions. Direct detection experiments are primarily sensitive to WIMP scattering on nuclei. Both signals are dominated by 
WIMP-quark interactions, so any comparison with the ILC bounds requires making theoretical assumptions regarding the relation between WIMP-quark and WIMP-electron couplings. Two distinct scenarios are plausible. First, the WIMPs may couple to quarks directly, at tree level, through operators similar to those in Eq.~\leqn{ops}. Second, the WIMP may only couple directly to leptons, in which case the dominant coupling to quarks appears via the one-loop diagram shown in Fig.~\ref{fig:oneloop}. (Which of these scenarios is realized depends on the quantum numbers and couplings of the new particles at scale $\Lambda$ that induce the WIMP couplings to the SM.) Furthermore, while collider bounds are relatively insensitive to the structure of the operator, the bounds from direct detection vary widely depending on this structure. Let us discuss a few interesting possibilities.

{\bf Direct WIMP-Quark Couplings:} If the WIMP-quark operator mediates spin-independent WIMP-nucleon scattering at low velocities, such as scalar or vector operators, then the bounds from direct detection are very strong. The current XENON100 bound~\cite{XENON100} rules out operators of this kind up to $\Lambda\sim 20$ TeV. This is well above the ILC sensitivity even in the most optimistic scenario, but it is worth emphasizing two points. First, XENON100 (and most other direct detection experiments) rapidly lose sensitivity for WIMP masses below 10 GeV, and, as is already well known, collider searches provide the more sensitive probe in that region. Second, there is no {\it a priori} reason why the scale suppressing the quark and lepton operators should be the same. 

If the WIMP-quark operator only mediates spin-{\it dependent} WIMP-nucleon scattering ({\it e.g.} axial-vector or tensor couplings), then the direct detection bounds on $\Lambda$ are much weaker, of order a few hundred GeV. ILC searches will be sensitive to much higher scales, and in addition will cover the region of low-mass WIMPs. 

For all operator structures, the current LHC bounds on $\Lambda$ are of order 700 GeV, well below the ILC sensitivities. 
(Although, once again, we emphasize that there is no {\it a priori} reason why the scale suppressing the quark and lepton operators should be the same.)

{\bf One-loop WIMP-Quark Coupling Only:} If no tree-level WIMP-quark operators are present, the dominant contribution to WIMP-quark scattering appears at one loop, see Fig.~\ref{fig:oneloop}. The LHC bounds are irrelevant in this situation due to the weakness of the interaction, so let us focus on direct detection. Since the momentum exchange in direct detection is well below $\Lambda_{\rm QCD}$, the $Z$ boson contribution can be neglected, and the photon couples to the nucleon as a whole, not to individual partons. Thus, only protons participate in this interaction. For simplicity, let us further assume that the WIMPs only couple to electrons (not $\mu$ or $\tau$), and that the coupling occurs through one of the operators listed in~\leqn{ops}. With these assumptions, it is straightforward to evaluate the cross section of elastic WIMP-proton scattering, $\sigma_p$, for given $M_\chi$ and $\Lambda$, see Ref.~\cite{LEPshines}. Note that the interaction in Fig.~\ref{fig:oneloop} is unavoidable, and adding tree-level WIMP-quark couplings can only increase the cross section unless there are accidental cancellations. Thus, our $\sigma_p$ can be thought of as an approximate model-independent lower bound on the direct detection cross section for given $M_\chi$ and $\Lambda$. 

\begin{figure}[h!]
\begin{center}
\centerline {
\includegraphics[width=3.3in]{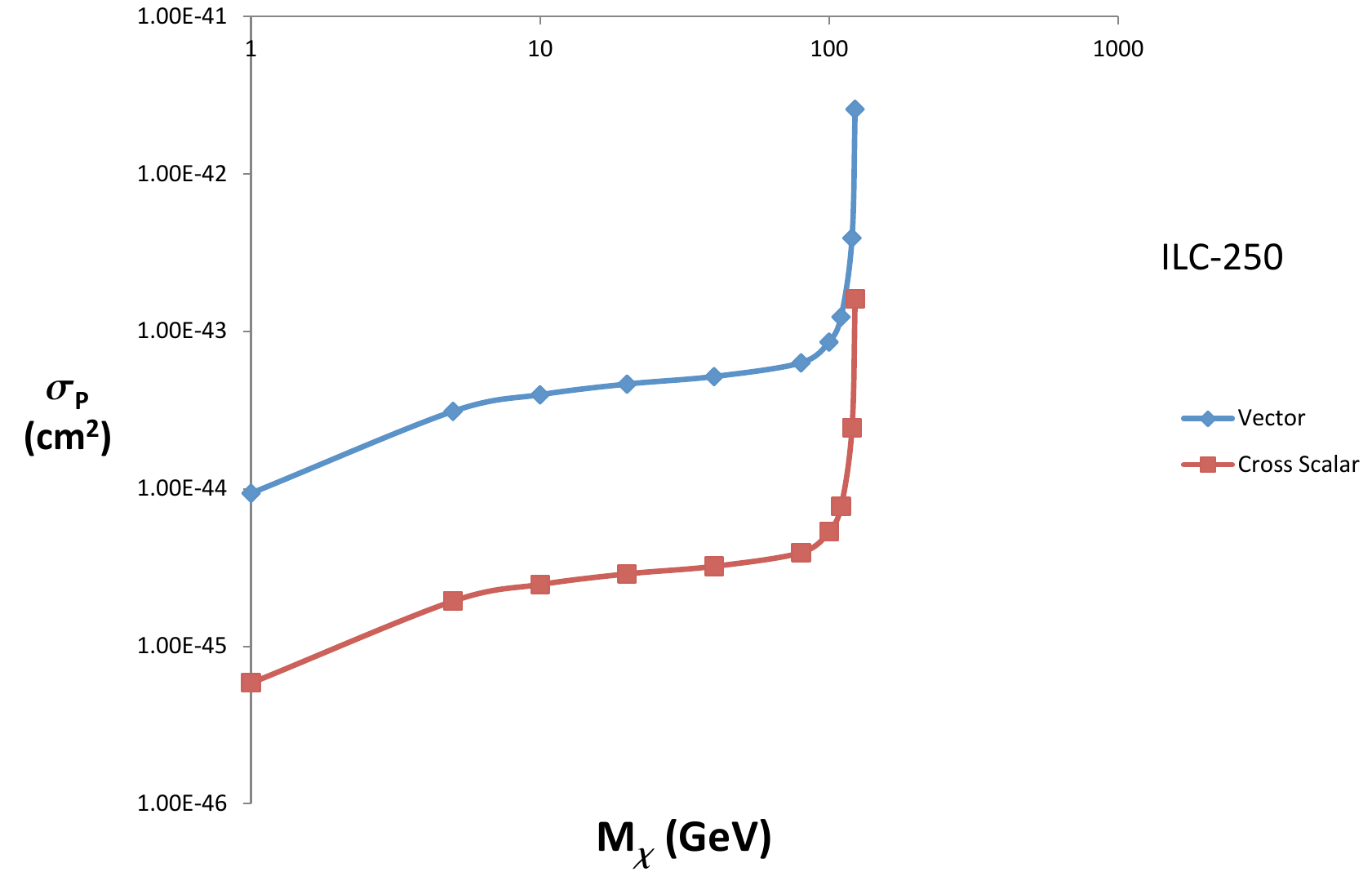}
\includegraphics[width=3.3in]{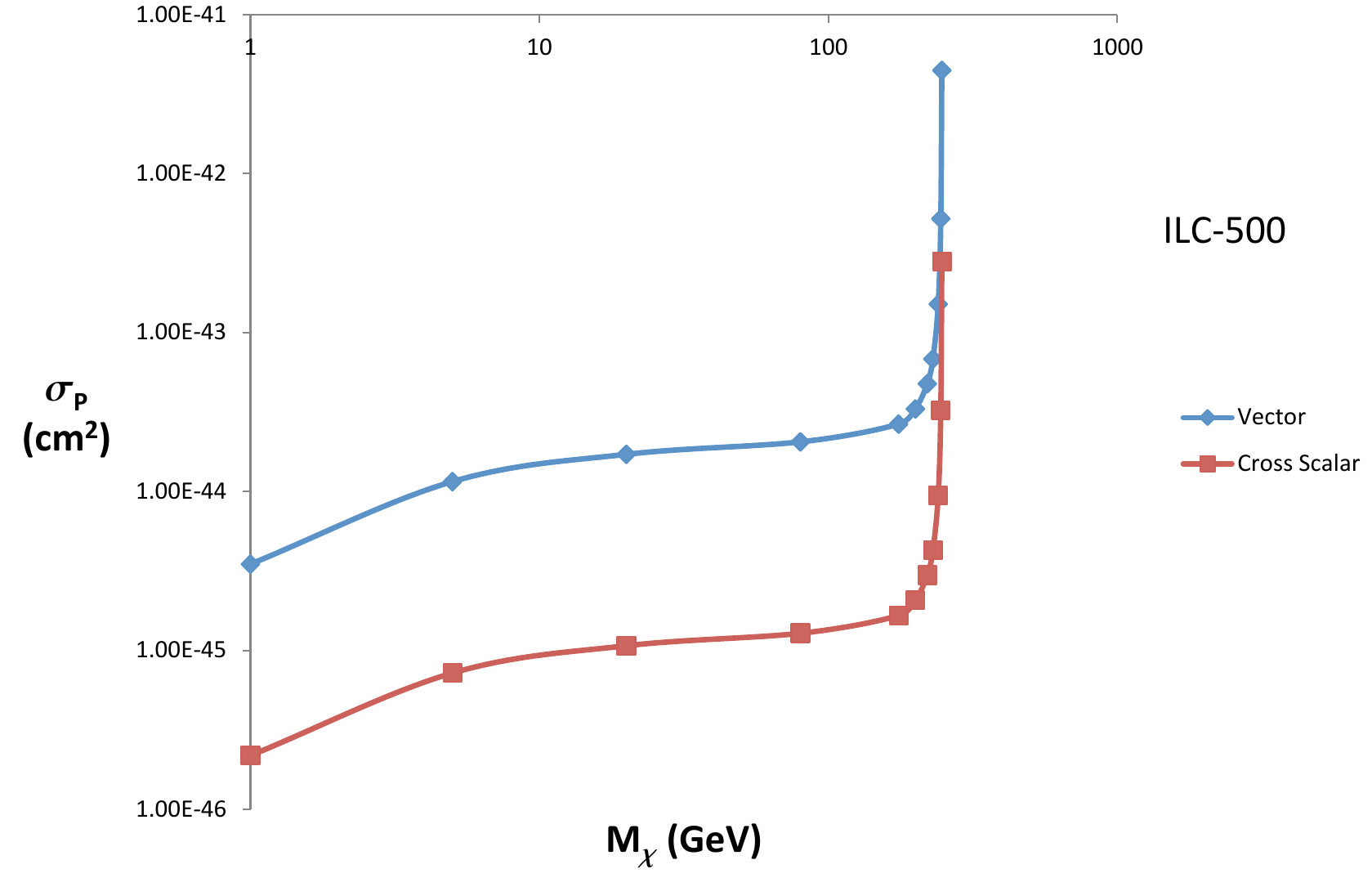}
}
\centerline {
\includegraphics[width=3.3in]{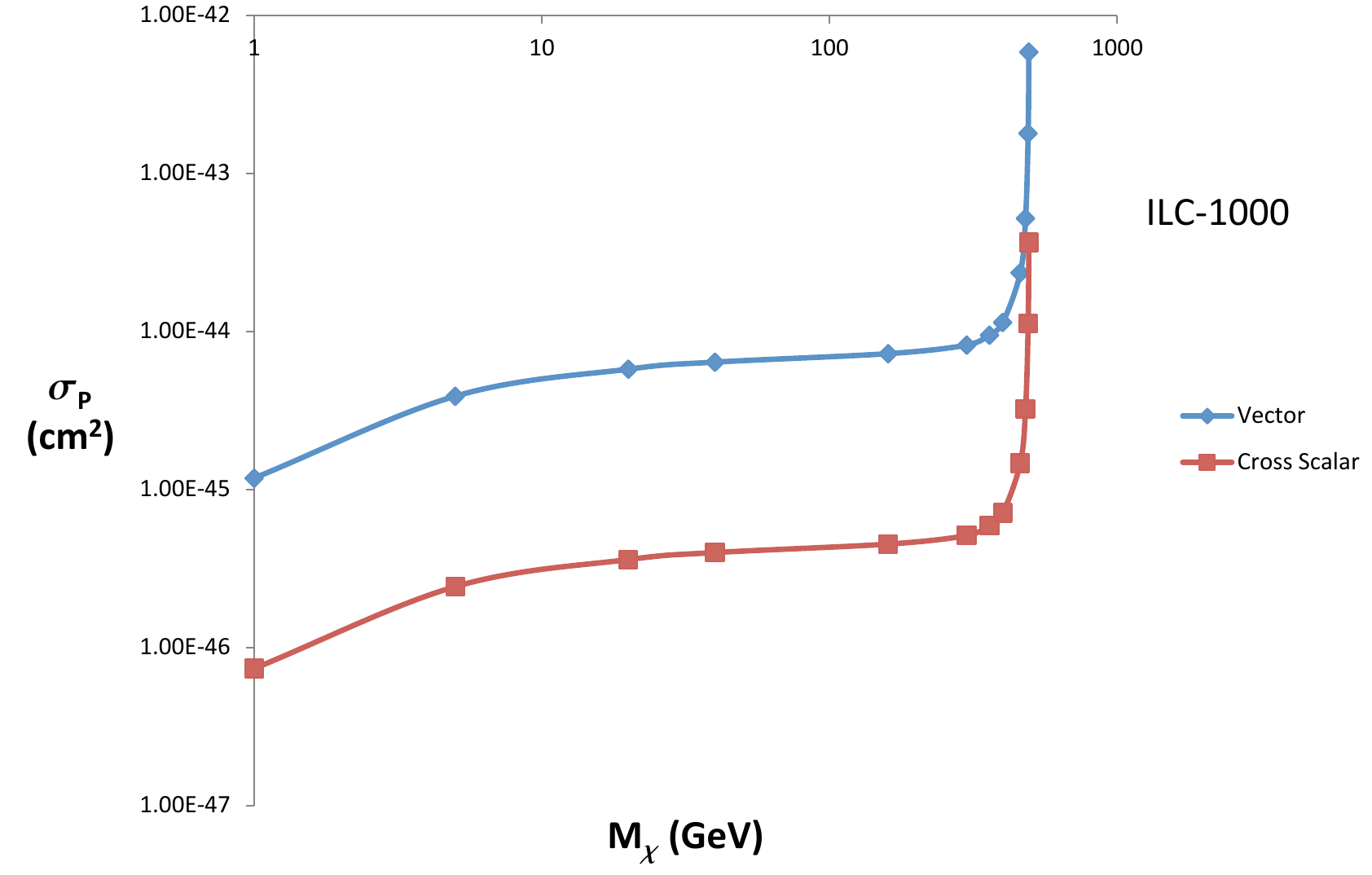}
\includegraphics[width=3.3in]{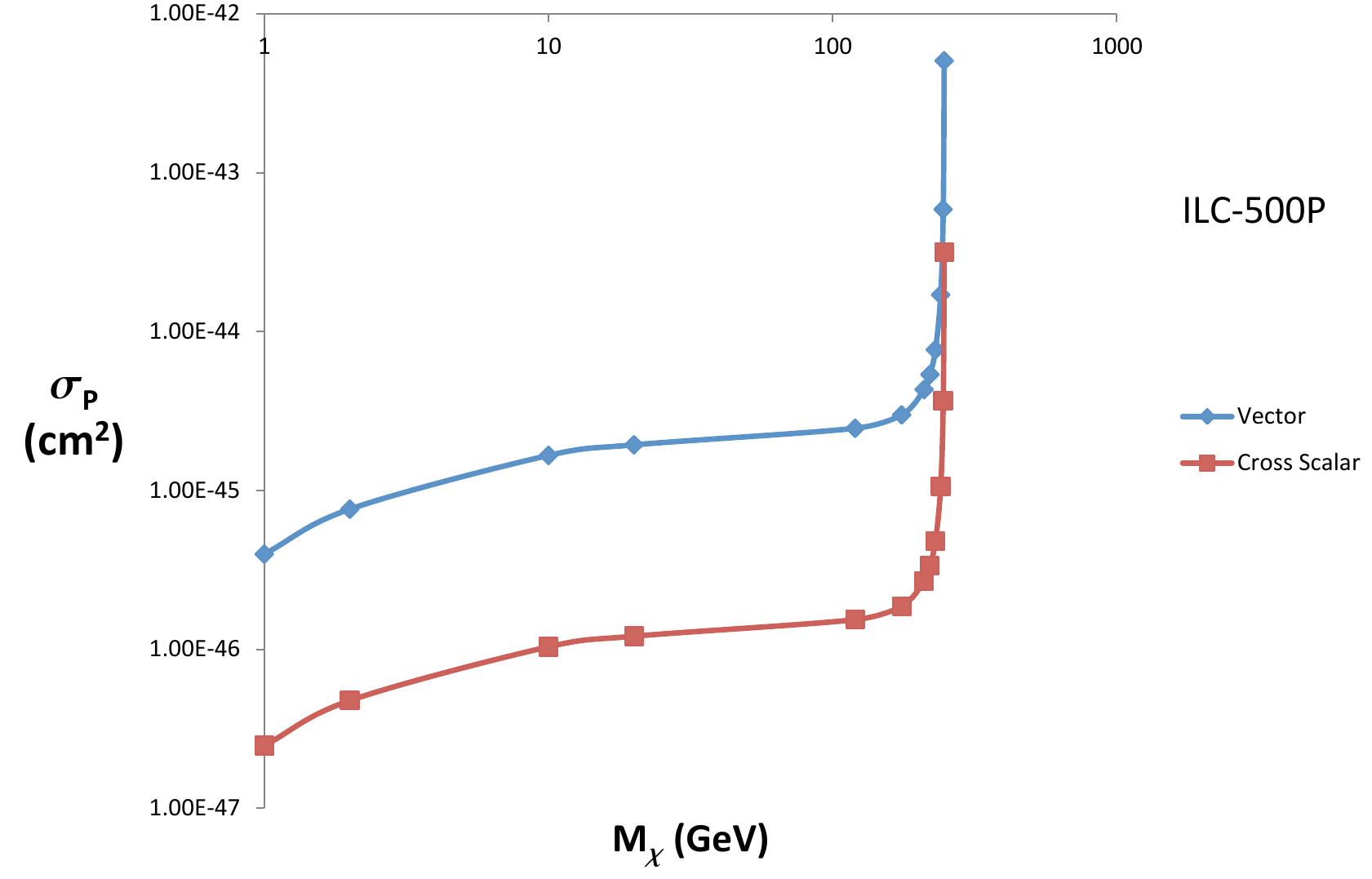}
}
\centerline {
\includegraphics[width=3.3in]{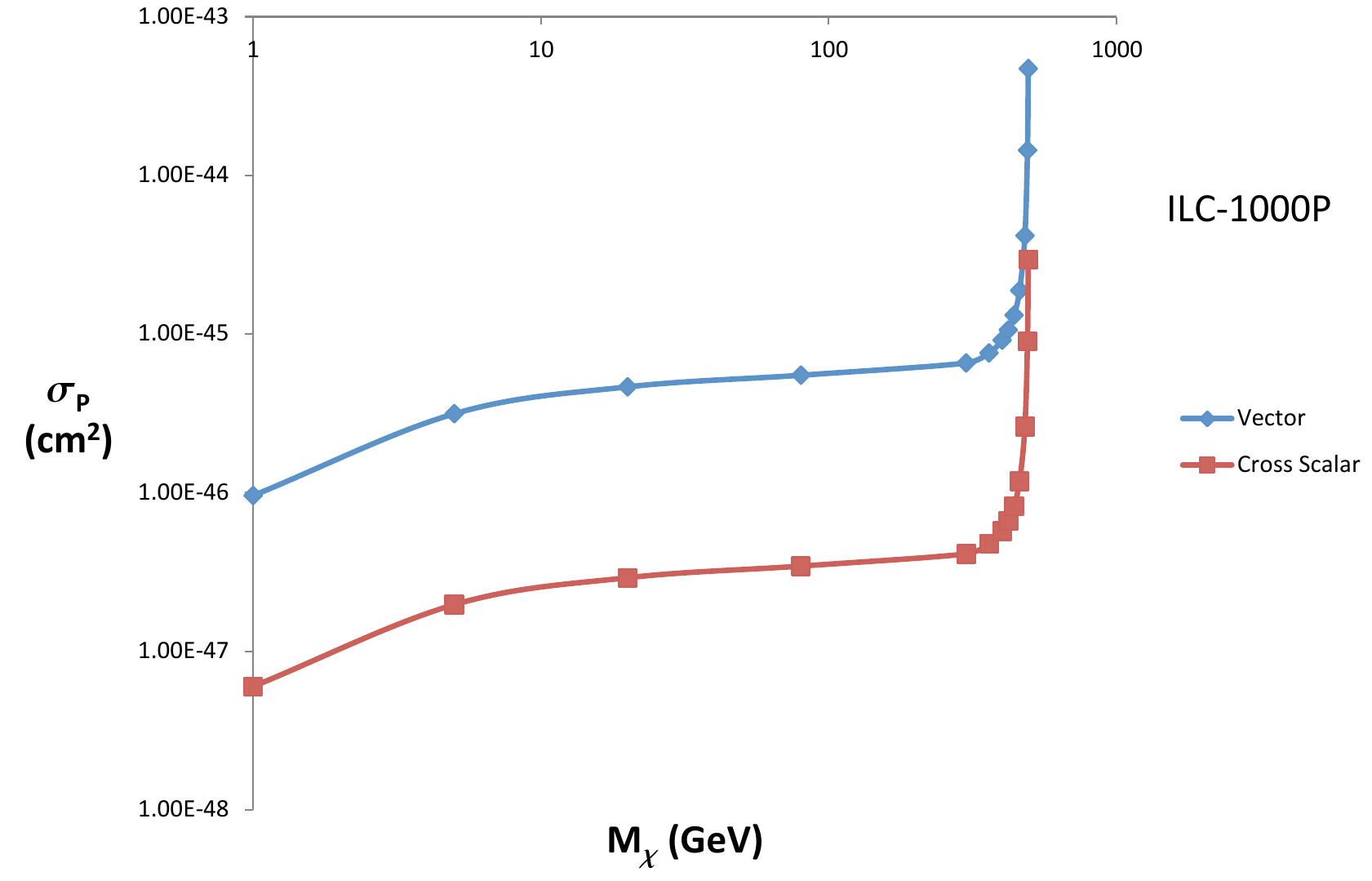}
}
\caption{Direct detection (WIMP-proton elastic scattering) cross section induced by the one-loop diagram in Fig.~\ref{fig:oneloop}, for values of $\Lambda$ at the outer edge of the ILC experimental sensitivity.}
\label{fig:direct}
\end{center}
\end{figure}

In Fig.~\ref{fig:direct}, we plot the contours of $\sigma_p$ corresponding to the ILC reach for various run scenarios. 
Of the four operators considered in this paper, only the vector and $t$-channel scalar couplings give non-zero contribution to the diagram in Fig.~\ref{fig:oneloop}. The other two operators would produce much smaller direct detection cross sections for the same $\Lambda$, and are not included in the plots. To compare with the published experimental bounds on WIMP-nucleon cross sections, one needs to rescale those bounds to take into account that in our model only protons, and not neutrons, couple to WIMPs, and the cross section depends on the momentum exchange. Including these effects, the current XENON100 bound rules out $\sigma_p$ down to about $10^{-43}$ cm$^2$ for masses above 50 GeV or so, with sensitivity decreasing rapidly for lower masses~\cite{LEPshines}. This leaves a large region of allowed parameter space which can be explored by the ILC.

\section{Conclusions and Outlook}
\label{sec:conc}

In this paper, we analyzed the sensitivity of the proposed high-energy $e^+e^-$ collider, the ILC, for a signal of direct WIMP pair production, accompanied by a photon. We employed the operator formalism to describe the WIMP-electron coupling, giving the analysis a high degree of model-independence. We found that the ILC-250, the ``Higgs factory" stage currently under active discussion, can extend the reach in terms of $\Lambda$ (the scale where the WIMP-electron interaction is induced) by a factor of $2.5-3$ compared to the current best bounds from LEP-2. Higher energy ILC upgrades can yield further spectacular improvements in sensitivity, especially if a beam polarization option is available. In all cases, the ILC reach allows to probe WIMP annihilation cross sections well below the value required for the thermal relic density to match the observed value. In other words, the ILC can discover WIMPs even if the process $\chi\chi\to e^+e^-$ only contributes a tiny fraction of the total annihilation rate $\chi\chi\to$ SM in the early universe. We also discussed comparison with direct detection experiments and collider searches at the LHC. While no firm model-independent conclusions could be drawn, we argued that the ILC will be able to explore large regions of parameter space consistent with present bounds. 

If the $\gamma+\me$ signal is discovered, the next task will be to measure the WIMP mass, the scale $\Lambda$, and to characterize the helicity structure of the operator coupling WIMPs to electrons. It would be interesting to study quantitatively the potential of the ILC to make these measurements. (See Refs.~\cite{ListEtAl2},~\cite{KKMP} for similar studies in slightly different theoretical frameworks.) We leave this issue for future work.

\vskip0.8cm
\noindent{\large \bf Acknowledgments} 
\vskip0.3cm

We would like to thank Michael Saelim, Bibhushan Shakya and Yuhsin Tsai for useful discussions. This research is supported by the U.S. National Science Foundation through grant PHY-0757868 and CAREER grant PHY-0844667.

\vskip0.8cm
\noindent{\large \bf Note added:} 
\vskip0.3cm

While the writing of this paper was being completed, Ref.~\cite{Dreiner:2012xm} appeared which overlaps with our work.

\begin{appendix}

\section{Analytic Formulae for Differential Cross Sections}
\label{app:analytics}

In this Appendix, we list analytic expressions for double-differential cross sections of the process $e^+e^-\to\bar{\chi}\chi \gamma$, in photon energy $E_\gamma$ and its angle $\theta$ with respect to the electron beam, for each of the four WIMP-electron interaction operators listed in Eq.~\leqn{ops}. For convenience, we define 
\beq
z=\frac{2E_\gamma}{\sqrt{s}},~~~\mu=\frac{M_\chi}{\sqrt{s}}. 
\eeq{defs}
Kinematically accessible range of $z$ is $[0, 1-4\mu^2]$. The formulas below give the double-differential cross section, $\frac{d^2\sigma}{dz d\cos\theta}$, for polarized electrons and positrons: for example, ``LR" corresponds to the $e_L^- e_R^+$, etc. For the case of vector interaction, we obtain:
\beqa
{\rm LR=RL:}~&~&\frac{\alpha}{12\pi^2}\frac{s}{\Lambda^4}\frac{1}{z\sin^2\theta}\sqrt{\frac{1-z-4\mu^2}{1-z}} 
\left( 1-z+2\mu^2\right) \left[ 4 ( 1-z ) + z^2(1+\cos^2\theta) \right];\CR
{\rm RR=LL:}~&~&0\,.
\eeqa{Vunpol} 
The case of axial-vector interaction yields
\beqa
{\rm RR=LL:}~&~&\frac{\alpha}{12\pi^2}\frac{s}{\Lambda^4}\frac{1}{z\sin^2\theta}\left({\frac{1-z-4\mu^2}{1-z}} \right)^{3/2} \left( 1-z\right) \left[ 4 ( 1-z ) + z^2(1+\cos^2\theta) \right];\CR
{\rm LR=RL:}~&~&0\,.
\eeqa{AVunpol} 
For the ``$s$-channel" scalar interaction, we find:
\beqa
{\rm RR=LL:}~&~& \frac{\alpha}{8\pi^2}\frac{s}{\Lambda^4}\frac{1}{z\sin^2\theta}\left({\frac{1-z-4\mu^2}{1-z}} \right)^{3/2} \left( 1-z \right) \left( 2 ( 1-z ) + z^2 \right),\CR
{\rm LR=RL:}~&~&0\,;
\eeqa{SSunpol} 
while for the ``$t$-channel" scalar interaction case, 
\beqa
{\rm RR=LL:}~&~& \frac{\alpha}{192\pi^2} \frac{s}{\Lambda^4}\frac{1}{z\sin^2\theta}\sqrt{\frac{1-z-4\mu^2}{(1-z)^3}} 
\Bigl[ (1-z)\left( (2-z)(3z^2-6z+4) -\cos^2\theta \right) \CR & & -2\mu^2\left(\cos^2\theta+4(1-z)-z(3z^2-6z+4)\right) \Bigr],\CR
{\rm LR=RL:}~&~& \frac{\alpha}{96\pi^2}\frac{s}{\Lambda^4}\frac{1}{z\sin^2\theta}\sqrt{\frac{1-z-4\mu^2}{1-z}} \left( 1-z+2\mu^2\right) \left[ 4 ( 1-z ) + z^2(1+\cos^2\theta) \right]. \CR
\eeqa{STunpol} 
We have checked that the Ward identities (vanishing of amplitudes upon replacement of the photon polarization vector by its momentum) work for all cases. 

We denote the polarization of the electron and positron beams, respectively, by $P_-$ and $P_+$. In our notation, $-1\leq P_\pm\leq 1$; $P_-=+1$ corresponds to pure $e^-_R$ and $P_+=+1$ corresponds to pure $e^+_L$, while $P_\pm=0$ corresponds to unpolarized beams. With this notation, the cross section is given by
\beqa
\sigma &=& \left(\frac{1-P_-}{2}\right)\left(\frac{1-P_+}{2}\right)\,\sigma_{{\rm LR}} +\left(\frac{1+P_-}{2}\right)\left(\frac{1+P_+}{2}\right)\,\sigma_{{\rm RL}} \CR &+& \left(\frac{1-P_-}{2}\right)\left(\frac{1+P_+}{2}\right)\,\sigma_{{\rm LL}} +\left(\frac{1+P_-}{2}\right)\left(\frac{1-P_+}{2}\right)\,\sigma_{{\rm RR}}\,. 
\eeqa{xsec}

\end{appendix}

\end{document}